# Without journalists, there is no journalism: the social dimension of generative artificial intelligence in the media

**Simón Peña-Fernández; Koldobika Meso-Ayerdi; Ainara Larrondo-Ureta; Javier Díaz-Noci**

**Nota**: Este artículo se puede leer en español en:
*https://revista.profesionaldelainformacion.com/index.php/EPI/article/view/87329*

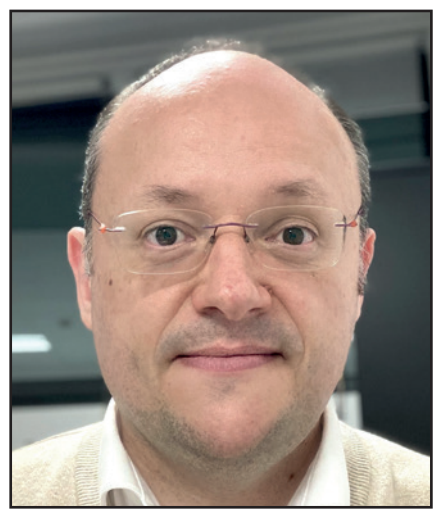

**Simón Peña-Fernández** ✉
*https://orcid.org/0000-0003-2080-3241*

*Universidad del País Vasco (UPV/EHU)*
*Facultad de Ciencias Sociales y de la Comunicación*
Barrio Sarriena, s/n
48940 Leioa (Vizcaya), Spain
*simon.pena@ehu.eus*

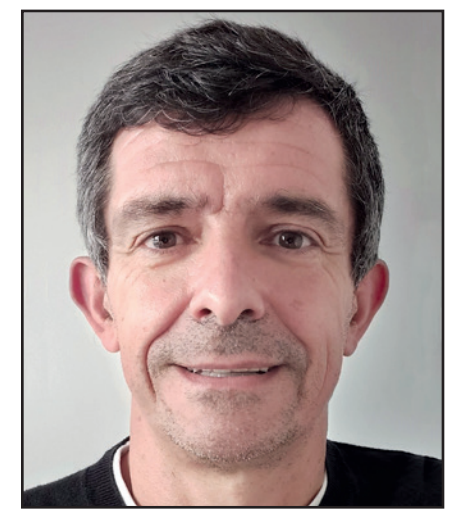

**Koldobika Meso-Ayerdi**
*https://orcid.org/0000-0002-0400-133X*

*Universidad del País Vasco (UPV/EHU)*
*Facultad de Ciencias Sociales y de la Comunicación*
Barrio Sarriena, s/n
48940 Leioa (Vizcaya), Spain
*koldo.meso@ehu.eus*

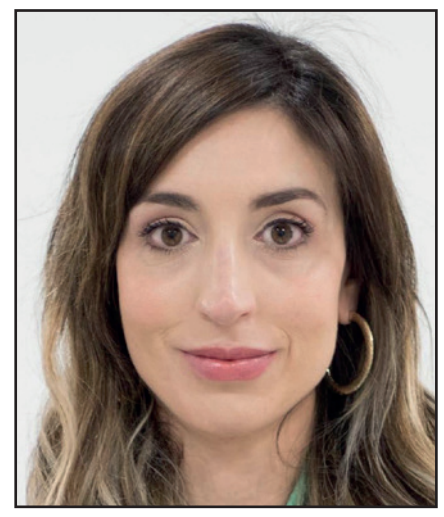

**Ainara Larrondo-Ureta**
*https://orcid.org/0000-0003-3303-4330*

*Universidad del País Vasco (UPV/EHU)*
*Facultad de Ciencias Sociales y de la Comunicación*
Barrio Sarriena, s/n
48940 Leioa (Vizcaya), Spain
*ainara.larrondo@ehu.eus*

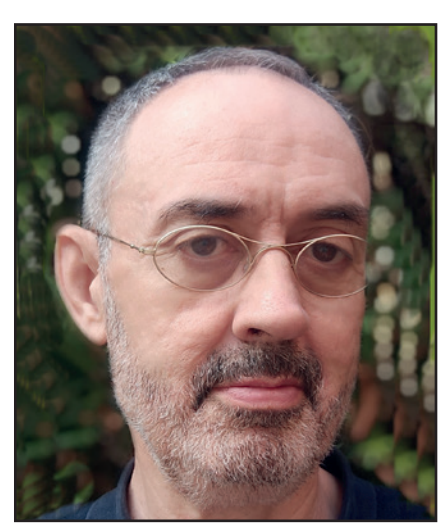

**Javier Díaz-Noci**
*https://orcid.org/0000-0001-9559-4283*

*Universitat Pompeu Fabra*
*Departament de Comunicació*
Roc Boronat, 138
08018 Barcelona, Spain
*javier.diaz@upf.edu*

**Abstract**

The implementation of artificial intelligence techniques and tools in the media will systematically and continuously alter their work and that of their professionals during the coming decades. To this end, this article carries out a systematic review of the research conducted on the implementation of AI in the media over the last two decades, particularly empirical research, to identify the main social and epistemological challenges posed by its adoption. For the media, increased dependence on technological platforms and the defense of their editorial independence will be the main challenges. Journalists, in turn, are torn between the perceived threat to their jobs and the loss of their symbolic capital as intermediaries between reality and audiences, and a liberation from routine tasks that subsequently allows them to produce higher quality content. Meanwhile, audiences do not seem to perceive a great difference in the quality and credibility of automated texts, although the ease with which texts are read still favors human authorship. In short, beyond technocentric or deterministic approaches, the use of AI in a specifically human field such as journalism requires a social approach in which the appropriation of innovations by audiences and the impact it has on them is one of the keys to its development. Therefore, the study of AI in the media should focus on analyzing how it can affect individuals and journalists, how it can be used for the proper purposes of the profession and social good, and how to close the gaps that its use can cause.

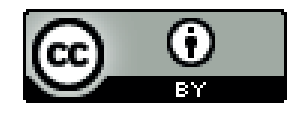

**Funding**

This work is part of the research project TED2021-130810B-C22 funded by *Agencia Estatal de Investigación* (*AEI*) of the Spanish *Ministry of Science and Innovation* (*Micin)* 10.13039/501100011033 and by the *European Commission* NextGeneration EU/PRTR. It is also part of the scientific production of the *Gureiker* consolidated research group of the *Basque University System* (IT1496-22).

## 1. Introduction

The advance of disruptive technologies that deepen the digitization of the media affects some of the issues that until now had been considered exclusively human. In addition, according to the analysis of the media's own discourse on artificial intelligence (AI), AI itself will be omnipresent and applicable to any type of task performed by people (**Brennen**; **Howard**; **Nielsen**, 2022).

Generally speaking, the indiscriminate use of the term "artificial intelligence" has led to its use with a wide variety of meanings (**Broussard** *et al.*, 2019), always referring to processes in which technology simulates human intelligence and that allows for computers and machines to behave in a similar way to people.

AI encompasses a complex set of techniques from the field of computer science that cut across all disciplines and social domains, and which can be further divided into seven sub-areas: machine learning, computer vision, speech recognition, natural language processing, automatic planning, expert systems, and robotics (**De-Lima-Santos**; **Ceron**, 2021).

Today, the development of AI in industrial processes is high on the research agenda at every level, and the anticipated potential benefits are not only limited to the economic and industrial dimensions, but also extend to the social sphere. Thus, the *European Commission*'s "White Paper on Artificial Intelligence" (2020) considers that its use can make it possible to meet challenges such as improving the quality of democracy or the provisioning of quality public services. The centrality of the service to citizens, and not the simple technological development of an industrial process, is one of the characteristics that should define this process so that it contributes to achieving sustainable economic and social growth (*Comisión Europea*, 2020).

Despite these promises, the growing implementation of AI also carries risks. The dark underside of its potential includes, among other dangers, increasing existing gaps -of gender, race, social class, etc.- through the use of biased data, problematic automated decisions, or intrusion into private life (**Brundage** *et al.*, 2018), owing, in part, to the use of partial or outdated data banks for model training or errors in their design, which can cause multiple ethical repercussions (**Dörr**; **Hollnbuchner**, 2017; **Ufarte-Ruiz**; **Calvo-Rubio**; **Murcia-Verdú**, 2021; **Barceló-Ugarte**; **Pérez-Tornero**; **Vila-Fumàs**, 2021; **Deuze**; **Beckett**, 2022), and which European institutions have also begun to address (*European Commission*, 2022a; 2022b).

If we focus on the specific field of the media, automation and AI have developed a multitude of applications in all phases of the information process (**Wu**; **Tandoc**; **Salmon**, 2019; **Marconi**, 2020; **Sánchez-García** *et al.*, 2023) through a varied range of resources, ranging from the use of algorithms for the analysis of consumption habits to the tracking of trends on social networks. This includes the development of tools to identify disinformation (**Ruffo**; **Semeraro**, 2022; **García-Marín**, 2022) or to moderate it in the comments section.

More specifically journalistic uses, on the contrary, are commonly associated with automation and generative AI, i.e., algorithmic processes that convert data into narrative texts and news, with limited or no human intervention beyond the initial programming process (**Carlson**, 2015). Ultimately, their use may involve a redefining of existing models from the communicative perspective of the media, since these technologies can be used as content-generating agents and not as mere mediators of human communication (**Guzman**; **Lewis**, 2020), which may even lead to synthetic media without journalists (**Ufarte-Ruiz**; **Murcia-Verdú**; **Túñez-López**, 2023).

Nor can we ignore the possible problems of job insecurity that the adoption of AI systems may cause for journalists (**López-Jiménez**; **Ouariachi**, 2020; **Stenbom**; **Wiggberg**; **Norlund**, 2021), already greatly accentuated after the successive economic crises since 2008 and other types of job inequality (**Díaz-Noci**, 2023).

Therefore, AI confronts journalism not only with challenges stemming from the adoption of a new technology in its ongoing digital transformation process, but also with issues that affect its intrinsically human nature. Thus, automated models pose new ontological models of relationship between humans and technology (**Primo**; **Zago**, 2015; **Lewis**; **Guzman**; **Schmidt**, 2019) in a profession traditionally characterized by the social and affective impact of the topics covered by the media, as well as of the interpersonal relationships established between journalists, sources, and audiences (**Riedl**, 2019).

In short, the application of AI in the media fuels the debate on how technological advances interrelate with the social dimension of journalism, which Kapuściński summarized in *Il cinico non è adatto a questo mestiere*:

> "I believe that, to practice journalism, above all, you have to be a good man, or a good woman: good human beings. Bad people cannot be good journalists. If you are a good person you can try to understand others, their intentions, their faith, their interests, their difficulties, their tragedies, and become immediately, from the very first moment, part of their destiny. This is a quality that in psychology is called empathy. Through empathy one can understand the character of one's interlocutor and naturally and sincerely share the fate and problems of others" ["*Credo che per fare del giornalismo si debba essere innanzi tutto degli uomini buoni, o delle donne buone: dei buoni esseri umani. Le persone cattive non possono essere dei bravi giornalisti. Se si è una buona persona si può tentare di capire gli altri, le loro intenzioni, la loro fede, i loro interessi, le loro difficoltà, le loro tragedie. E diventare immediatamente, fin dal primo momento, parte del loro destino. È una qualità che in psicologia viene chiamata "empatia". Attraverso l'empatia si può capire il carattere del proprio interlocutore e condividere in maniera naturale e sincera il destino e i problemi degli altri." (***Kapuscinski***, 2000)*

Indeed, the intimacy of journalists with the emotions and desires of their readers has been one of the most specifically human features of the profession that has been emphasized, as well as its nature as a tool for the development of fundamental rights. Thus, the *Federation of Associations of Journalists of Spain* (*FAPE*), as well as events such as World Press Freedom Day, have claimed during the last few years that "without journalists, there is no journalism" and that "without journalism there is no democracy" (*FAPE*, 2014), a democracy with respect to which the profession has a moral obligation (**Strömbäck**, 2005).

In this context, this article aims to identify the social dimension of the main challenges of journalism in the age of AI and algorithms in relation to three essential aspects: its business and industrial development, the impact on media professionals, and the effects on audiences.

## 2. AI and journalism

Content automation is not new in the media domain (**García-Orosa**; **Canavilhas**; **Vázquez-Herrero**, 2023); rather, it has been around for at least four decades, albeit with a limited scope thus far (**Lindén**, 2017).

The names it has been referred to within the broader framework offered by the practices of computational journalism and computer-assisted journalism (**Codina**; **Vállez**, 2018; **Parratt-Fernández**; **Mayoral-Sánchez**; **Mera-Fernández**, 2021) have been diverse: "machine-written journalism" (**Van-Dalen**, 2012), "algorithmic journalism" (**Anderson**, 2012), "robotic journalism" (**Clerwall**, 2014), or more commonly, "automated journalism" (**Graefe**, 2016; **Moran**; **Shaikh**, 2022).

In the case of media, talking about AI almost always means talking about automated content, although in reality its applications (**Chan-Olmsted**, 2019), as well as its theoretical developments, have been highly varied, particularly during the last decade (**Parratt-Fernández**; **Mayoral-Sánchez**; **Mera-Fernández**, 2021).

Thus, various media, such as *The New York Times*, *The Washington Post*, or *Le Monde*, and news agencies such as *Reuters* or *Associated Press* (**Fanta**, 2017; **Túñez-López**; **Toural-Bran**; **Cacheiro-Requeijo**, 2018; **Chan-Olmsted**, 2019), have developed initiatives for automated content production, usually in collaboration with technology companies (**Dörr**, 2016; **Lindén**; **Tuulonen**, 2019). These projects have been based on automated text planning, whereby information is created by connecting predetermined templates to databases (**Carlson**, 2015; **Biswal**; **Gouda**, 2020), resulting in texts with more or less repetitive structures (**DalBern**; **Jurno**, 2021).

The ideal areas for the development of this automated content so far have been topics such as weather, financial, or sports information (**Canavilhas**, 2022), for which there are highly structured databases that allow for efficient information extraction to be easily automated (**Graefe**; **Bohlken**, 2020), and where speed prevails over depth in the analysis (**Kim**; **Kim**, 2017). The development of chatbots can also be included in this area (**Lokot**; **Diakopoulos**, 2016; **Jones**; **Jones**, 2019; **Veglis**; **Maniou**, 2019), although their presence and influence in the media have been more limited.

However, the evolution of content automation has gone a step further with the popularization of generative AI systems (for example, *ChatGPT*, *Dall-E*, or *Midjourney*), and various media outlets have openly reported the use of AI in the production of news material on a systematic basis. The debate regarding the real scope of the adoption of this technology -not new, but with a degree of development and implementation never seen before- and whether it is an opportunity or a threat to journalism (**Adami**, 2023), has thus been rekindled.

## 3. Methodology

For this article, a systematized literature review (**Codina**, 2020) was carried out on the basis of the concepts of "automated journalism" and "robot journalism," as well as their functional synonyms and other related definitions (algorithms, artificial intelligence, etc.) in the field of journalism and media, in both Spanish and English, which complements some previous reviews (**Calvo-Rubio**; **Ufarte-Ruiz**, 2021; **García-Orosa**; **Canavilhas**; **Vázquez-Herrero**, 2023). The period from 2000 to the present was established as the time range. The search was carried out using the main academic databases (*Web of Science*, *Scopus*) and completed using queries in the *Dialnet* directory and the *Google Scholar* search engine.

The results obtained were analyzed and categorized on the basis of quantitative data –the number of references obtained– as well as qualitative data –the topics dealt with on the basis of summaries and keywords– giving priority to empiri-cal studies dealing with the social dimension of AI, particularly in its generative modality, and the opinions of audiences, professionals, and media managers. Articles dealing exclusively with technical aspects or nonjournalistic uses of AI were excluded. This selection process resulted in a final sample of 223 texts. The resulting texts were analyzed, evaluated, and summarized (**Grant**; **Booth**, 2009) in terms of the impact of AI on the three main subjects of the communicative process: the media, its professionals, and its audiences.

## 4. Results

### 4.1. Industrial dimension of news production

Twenty-five years after the emergence of the first cybermedia, the development of digital technologies continues to transform journalistic and news content, a systemic change that has affected the media equally in all the facets of their activity. It has accelerated and expanded the area of information dissemination, transformed its business models, and rethought the relationships between the different actors in the communication process (**Peña-Fernández**; **Lazkano-Arrillaga**; **García-González**, 2016).

Immersed in the paradigm of the fourth industrial revolution (**Micó**; **Casero-Ripollés**; **García-Orosa**, 2022), information companies advance in the improvement of their production processes in a complex hybrid media system (**Chadwick**, 2013) in which multiple actors create and disseminate content under the growing weight and power of the large technological giants (*Google*, *Meta*, etc.) (**Nielsen**; **Ganter**, 2022).

In this context, the main concerns for media managers in relation to content automation are its commercial viability and the way in which the investment needed to develop it can be acquired, as well as the acceptance that this type of content will have among their readers (**Dörr**, 2015; **Kim**; **Kim**, 2017).

In the face of this new development, **Boczkowski** (2004) points out that it should be taken into account that the media's attitude toward digital transformation has traditionally been reactive, defensive, and pragmatic. In other words, faced with the emergence of technological innovations, the media have imitated what their competitors were doing, protecting themselves from the initiatives of the large telecommunications companies and concerning themselves more with short-term threats than with long-term opportunities.

Some studies published to date have attributed the difficulty of implementing AI in the media to factors linked to investment costs, both for the development of proprietary applications and for the purchase of external tools. The cost of AI development being very high, even for large media (**Broussard** *et al*. 2019), and that not all of them need to develop it to meet their objectives, are also factors.

Similarly, the habitual caution and insularity that usually characterize the implementation of digital transformations in the media (**Beckett**, 2019) are reinforced by the doubts generated around the legal responsibility for the publication of content whose creation is not controlled at all ends, by the diffuse attribution of authorship, or the protection of intellectual property rights (**Montal**; **Reich**, 2017; **Lewis**; **Sanders**; **Carmody**, 2018; **Díaz-Noci**, 2020).

For the media, generative AI enters fully into the struggle for authorship and copyright, where human intervention and content creation are the main assets of journalistic companies in their fight against large platforms in the business of information distribution (**Díaz-Noci**, 2020).

In a profession whose main concern should be citizens (**Lemelshtrich**; **Nordfors**, 2009), content created in an automated way or through generative AI also poses relevant challenges relating to issues such as the creation of content about people, ethics in the use of data, or the transparency of algorithms (**Hansen** *et al.*, 2017; **Riedl**, 2019; **Ventura-Pocino**, 2021; **Pihlajarinne**; **Alén-Savikko**, 2022). These are in addition to those from previous digital transformation processes, such as the risk of polarization or limitations to pluralism that may result from the personalization of content (**Masip**; **Suau**; **Ruiz-Caballero**, 2020).

Finally, at a time when content created wholly or partially through generative AI is approaching exponential growth, expectation management will play a relevant role in the media, as expectations influence the way in which technological developments are perceived by the public and set research priorities that can influence their development and design (**Brennen**; **Howard**; **Nielsen**, 2022). Expectations have a performative character, as they can provide legitimization of technologies even before their success is proven, offer heuristic guidance that can help developers choose a path among existing ones, and provide coordination, mobilizing people and resources to build, design, and extend technologies (**Van-Lente**, 2012).

The management of expectations and promises therefore still has a central role to play in an ecosystem in which automated content creation in the media has been limited and in which there are still no widespread experiences of using generative AI for journalistic content creation.

### 4.2. Impact on professionals

In an ecosystem in constant transformation, this is not the first time that digital technological development has confronted journalists with the loss of their symbolic capital as mediators between reality and citizens. Without having to go too far back in time, at the dawn of Web 2.0 the rise of citizen journalism already opened the door to an eventual interactive and connective production in which users and the media coexisted and collaborated, as well as competed, in a joint construction of reality (**Deuze**, 2009). However, despite the existence of isolated "acts of journalism" linked to major events and catastrophes (**Holt**; **Karlsson**, 2014), it soon became clear that audiences were not interested in the sustained creation of journalistic content that competed with the media (**Masip**; **Ruiz-Caballero**; **Suau**, 2019).

The technological development of digital media, far from weakening the corporate sentiment of professionals, has contributed to the strengthening of their identity (**Ferrucci**; **Vos**, 2017). Journalists perceive themselves as an autonomous, self-regulating group (**Andersson**; **Wiik**, 2013) that fulfills a public service function in an impartial and neutral way (**Deuze**, 2005), belongs to organizations that share common goals (**Örnebring**, 2013) and whose task is the production of primary information (**Vos**; **Ferrucci**, 2018), and who consider veracity, contrast, and plurality of sources, as well as the distinction between facts and opinions, as some of the distinctive features of their profession (**Suárez-Villegas**, 2017).

AI once again presents professionals with the eventual loss of part of this symbolic capital and a social questioning that adds to the threat of job loss anticipated by publishers themselves (**Kim**; **Kim**, 2017), aggravated by the latest generative applications (**Elondou** *et al.*, 2023).

Thus, it is not surprising that professionals consider AI and automated content as jeopardizing the integrity of the profession (**Pérez-Dasilva** *et al.*, 2021; **Noain-Sánchez**, 2022). This has perhaps been contributed to, at least in part, by the depiction of replacement raised by the recurrent anthropomorphic image that often illustrates the emergence of "robot journalists" (**Lindén**, 2017; **DalBen**; **Jurno**, 2021), or the limited role attributed to communication professionals in a process conceived from a technocentric vision (**Carlson**, 2014; **Dörr**, 2015).

The importance of professionals' attitudes is not irrelevant, since the digital transformation has shown, among other issues, that the adaptation of innovations in newsrooms, in addition to industrial and economic challenges, has also had to face professionals' lack of training or their resistance to change (**Paulussen**, 2016; **Noain-Sánchez**, 2022). For these reasons, it is important to anticipate not only how technology will alter professional practice, but also the way in which this change is imagined by journalists themselves (**Gynnild**, 2014; **De-Haan** *et al.*, 2022; **Soto-Sanfiel** *et al.*, 2022).

On the contrary, theoretical approaches insist on approaching AI in the media as a set of tools and technologies that can free journalists from performing simple and repetitive tasks, which will allow them to spend more time on tasks that cannot be automated (**Van-Dalen**, 2012; **Young**; **Hermida**; 2015; **Wu**; **Tandoc**; **Salmon**, 2019; **DalBen**; **Jurno**, 2021). This liberation could allow them to add greater complexity and significance to their texts (**Brennen**, 2018) and devote greater dedication to research (**Flew** *et al.*, 2012; **Stray**, 2019) while helping them to overcome some of the current challenges of the profession, such as lack of information, decline in credibility, or crisis in business models (**Ali**; **Hassoun**, 2019). Ultimately, AI could facilitate the return of journalists to the essence of their profession, overcoming the post-Fordist model that limited their role to that of transcribers of facts (**Noain-Sánchez**, 2022).

In the face of deterministic approaches of a more dramatic nature, the challenge for journalists is not to avoid being replaced by disruptive technology, but to discern the way in which the ethical and normative values of the profession can be programmed into the automated generated content, and to find a way in which that content can be integrated into their work (**Diakopoulos**, 2019).

Even the most optimistic statements predict the creation of new jobs in the media associated with the emergence of novel professional profiles, such as the supervision of generated content (**Diakopoulos**, 2019). From this perspective, changes in news production technologies will not simply modify journalistic practices but also introduce what could be considered technologically specific ways of working (**Powers**, 2012), which will require greater collaboration of journalists with technical staff (**De-Lara**; **García-Avilés**; **Arias**, 2022).

This coexistence with a new technology that conditions and complements their work makes it necessary to rethink the professional roles and training needs of journalists. Its evolution will probably be oriented toward more hybrid modes of work (**Deuze**; **Beckett**, 2022), for which it will be necessary to offer new training profiles (**Calvo-Rubio**; **Ufarte-Ruiz**, 2020) in which specifically human traits of the profession, such as curiosity, skepticism, and critical thinking (**Thurman**; **Dörr**; **Kunert**, 2017), will also be deepened.

In any case, journalists show concern about the impact of AI on citizens (**De-Lara**; **García-Avilés**; **Arias**, 2022), and overwhelmingly express their desire to maintain control at all stages of news production (**Wu**; **Tandoc**; **Salmon**, 2019). The ultimate goal of the adoption of AI would be to reinforce the keys to journalistic work, such as creativity, listening, or sourcing (**Guzman**; **Lewis**, 2019).

In the eventual dispute between a competitive relationship and a complementary relationship with AI, journalists do not hesitate to embrace the latter, built on the defense of the cardinal values of the profession.

### 4.3. A human-centered media AI

However, in the development of automated content, as in all digital transformation processes, audiences will also play a key role. The fluctuation between the more technologically deterministic approaches, which predict a series of automatic effects -whether negative or positive- through the mere provision of technology, and the more constructivist and integrated approaches, which lead to the way in which the media and its professionals interact with such technologies, cannot refute that any technology that aspires to become an effective innovation requires a process of social appropriation (**Peña-Fernández**; **Lazkano-Arrillaga**; **Larrondo-Ureta**, 2019). Therefore, it is relevant to remember that technological development will not be the only thing driving the implementation of AI in the media (**Van-Dalen**, 2012); rather, social factors such as the acceptance of this content and the way in which its consumption becomes naturalized will also undoubtedly constitute one of the keys to its success.

Existing studies, predating more sophisticated generative AI practices, indicate, on the one hand, that formal emulation is perceived as efficient and that audiences do not find it easy to distinguish content created in an automated way (**Clerwall**, 2014; **Haim**; **Graefe**, 2017; **Graefe** *et al.*, 2018).

On the other hand, the success of the formal emulation that automated content achieves in terms of journalistic standards does not necessarily imply that audiences attribute the same functional authority to such content, since journalism provides not only information but also a way of knowing the world that has accumulated the epistemic authority necessary to be considered legitimate (**Carlson**, 2015).

Therefore, during the last decade, several empirical studies have evaluated automated content according to the three basic criteria for evaluating journalistic texts: the credibility of the source and text, the quality of the information, and the ease and amenity of reading (**Sundar**, 1999), and concluded that the objective style of journalism is a good refuge for automated content. Influenced by the way they are programmed, this content is perceived as more descriptive, objective, and informative (**Clerwall**, 2014; **Lui**; **Wei**, 2019), and is given at least equal credibility to that written by journalists (**Haim**; **Graefe**, 2017; **Melin** *et al.*, 2018; **Graefe** *et al.*, 2018; **Wölker**; **Powell**, 2018; **Zheng**; **Zhong**; **Yang**, 2018; **Liu**; **Wei**, 2019).

For readers, automated information written in an objective style and strictly following the rules of journalistic writing is virtually indistinguishable from that written by journalists. Moreover, the identification of nonhuman authorship does not reduce the credibility of the source itself nor of the content in this type of texts (**Tandoc**; **Yao**; **Wu**, 2020), most likely because of the commitment to impartiality attributed to their technical nature (**Gillespie**, 2014). Some studies even claim that the reception of information is more favorable if its authorship is identified as automated (**Jung** *et al.*, 2017), although such results are not unanimous (**Wadell**, 2018; 2019).

The different perception that readers (**Joris** *et al.*, 2021) have of the content by itself or in knowing its authorship -automated or human- introduces the relevance of the expectations it generates in the acceptance of the content. While the differences in reception are practically nonexistent in the case of more routine information, in the case of nonobjective texts, trust in automated content suffers, and the credibility attributed to them is lower than if they had been written by a journalist. This is probably influenced by "machine heuristics" (**Sundar**, 2008), whereby algorithms are credited with the ability to handle data objectively but are also expected to present results in the same way. Even readers' preconceived image of the depiction of robots in popular culture seems to affect the acceptance of automated content (**Sundar**; **Waddell**; **Jung** 2016).

The interpretation of these studies, however, is not unanimous and is nuanced. In their meta-analysis of research on the impact of automated content on audiences, **Graefe** and **Bohlken** (2020) state that the discrepancies are due to the experimental or descriptive nature of the studies, as the former tend to offer a somewhat more favorable view of human authorship. In any case, all the studies show a high degree of consensus in establishing very small differences in the credibility of automated information and somewhat greater differences in the perception of its quality compared with information produced by journalists.

Faced with similar credibility and a small differential perception of quality, the biggest differences between automated texts and those produced by journalists emerge in the ease with which the information is read. When authored by humans, the information is considered more engaging, enjoyable, and pleasant to read (**Clerwall**, 2014; **Graefe** *et al.*, 2018; **Melin** *et al.*, 2018; **Zheng**; **Zhong**; **Yang**, 2018), as well as eliciting greater emotional involvement than that which is generated in an automated fashion (**Lui**; **Wei**, 2019). Outside the realm of routine news based on objective data, interest in reading automated texts is diminished.

While the evolution of generative AI is expected to narrow this gap in content acceptance, one of the keys will be whether journalists should base their work on repeating existing models that AI already manages to replicate effectively, looking for differentiated styles, or recreating the generated texts on an automated basis.

## 5. Conclusions and discussion

Before the resounding emergence of generative AI, multiple technical processes developed from this field of computer science have been quietly making their way into media routines during the last five decades (**Beckett**, 2019; **Cools**; **Van-Gorp**; **Opgenhaffen**, 2022), both in support processes and in the automation of editorial content.

Within this general framework of digitization, the application of AI in industrial processes, including in the media, is developing as a top-down process in which institutional priorities coincide with developments led by large technology pla-tforms. This is an industry-led debate aimed at developing new business models and optimizing existing ones (**Lee** *et al.*, 2019), and which causes discourses to focus more on the capabilities of AI than on the consequences of its adaptation.

However, particularly in fields with a specifically human dimension such as journalism, it is relevant to underscore that AI is not only a technological product, but also a cultural one (**Guzman**; **Lewis**, 2020), and that no technological innovation -no matter how sophisticated- can move us away from our human nature (**Broussard**, 2019). As the father of cybernetics Norbert Wiener stated at the dawn of automation, the key is to consider the machine not as an end in itself, but as "a means to satisfy the demands of man, as a part of the human-mechanical system" (**Guéroult** *et al.*, 1966). Therefore, the first step in the application of AI is understanding it as a set of tools developed by humans in the service of human means and ends (**Broussard** *et al.*, 2019).

Thus, along with the predicted positive effects that can be attributed to its implementation, it is worth emphasizing that AI is a dual-use technology (**Brennen**; **Howard**; **Nielsen**, 2018) that allows for content to be generated in equal measure for malicious purposes, producing new vulnerabilities and risks (**Brundage**, 2018; **Karnouskos**, 2020). In addition, to the extent that they reproduce the real-life patterns in which they have been programmed, algorithms have biases that have been widely studied (**Pihlajarinne**; **Alén-Savikko**, 2022) and may contribute to reinforcing existing social gaps (**Eubanks**, 2017). As **Shilton** (2018) points out, the use of algorithms is deeply political, as they exude the values that their designers and developers have incorporated into them.

The focus in the debate on the application of AI to a specifically human activity such as journalism is therefore not on what the technology is capable of doing, but rather on how it can contribute to achieving its professional and social goals.

First, for the media and the companies that support them, AI can be a tool that contributes to increasing the productive efficiency of tasks that have a high human and organizational cost (**Thurman**; **Lewis**; **Kunert**, 2019). However, the costly implementation of these processes and their development once again stirs the debate about their increasing dependence on large technological platforms (**Danzon-Chambaud**, 2021; **Nielsen**; **Ganter**, 2022; **Simon**, 2022).

Despite new business opportunities (**Lindén**; **Tuulonen**, 2019), such as personalized content (**Møller**, 2022), the high cost of developing these applications and the limited capacity of the media to develop their own tools complicate their position in an ecosystem in which the difference in the technical potential of large and small players is growing. Therefore, one of the decisive challenges for the media will be in managing to enact their own organizational, institutional, and professional values in order not to remain in the hands of large technological platforms (**Diakopoulos**, 2019) in an industry that had found its specific niche in the digital ecosystem precisely with the creation of content. In the face of the impact of AI, one of the challenges will therefore be how the media manages to protect its editorial independence (**Lin**; **Lewis**, 2022; **Van-Drunen**; **Fechner**, 2022).

Early analyses indicate that the implementation of generative AI in media will probably not be as broad and deep as advocated by those who embrace the most enthusiastic and technologically deterministic predictions, but it will certainly offer valuable and relevant examples (**Brennen**; **Howard**; **Nielsen**, 2022). In their implementation, the complex and sometimes contradictory assimilation processes of the novel technical capabilities by the actors in the established social and material structures must be taken into account (**Boczkowski**, 2004). The social appropriation demonstrated through their effective use (**Echeverría**, 2008) will be the indicator of the extent to which they constitute an innovation for the media, a sector whose technological evolution traditionally occurs in a cumulative and nondisruptive way, resorting to external digital applications and systems (**De-Lara** *et al.*, 2015).

It will also be of interest to follow how the masthead under which content is published influences the perception of the quality of automated news (**Liu**; **Wei**, 2019) and therefore the institutional authority attributed to them, as well as the measures taken by the media to ensure transparency in their use.

Second, for journalists and professionals, the implementation of generative AI will again shake up the traditional dispute between editorial and business values, or in other words, the struggle between the commercial and journalistic soul of the media (**Andersson**; **Wiik**, 2013).

In the human appropriation of AI, professionals are the group most reluctant to apply it because of the professional and social undermining they perceive from its eventual widespread implementation (**Kim**; **Kim**, 2018). In any case, despite how AI may eventually constitute more than a simple channel and occupy a role generally attributed to people (**Lewis**; **Guzman**; **Smith**, 2019), this fear of replacement can be relegated to its conception as a set of tools and techniques at the service of journalism.

On the one hand, existing studies show that automated content is already competitive in routine matters, at least in certain highly tasked creation circumstances (**Haim**; **Graefe**, 2017). The automation of part of their tasks, in which human intervention will continue to occupy a relevant role (**Rojas-Torrijos**, 2019), may also be an opportunity to increase the cognitive value of journalistic work (**Túñez-López**; **Toural-Bran**; **Valdiviezo-Abad**, 2019) and escape from more routine and repetitive approaches (**Beckett**, 2019; **Graefe**; **Bohlken**, 2020).

AI provides speed, the ability to manage large volumes of information, verification tools, and personalized and multilingual content, and has already been gradually adopted in multiple news creation and distribution processes (**Wu**; **Tandoc**; **Salmon**, 2019). Faced with the risk of job loss, the entrenched professional ideology (**Deuze**, 2005) and corporate conceptions may have a mitigating effect on the direct impact of technology application (**Lindén**, 2017).

The eventual relief that the application of generative AI would bring to the creation of content of low added value or repetitive nature can put the focus back on more qualitative -and also more intrinsically human- aspects, such as the search for information, the interpretation of facts, creativity, humor, or criticism, i.e., issues that can contribute to improving their work. Journalists do not just write, they think, and journalism is more about asking the right questions than simply writing answers.

Therefore, hybrid or collaborative approaches between AI and journalists are reinforced (**Wadell**, 2018; 2019; **Wu**; **Tandoc**; **Salmon**, 2019; **Tejedor**; **Vila**, 2021), i.e., an integrative or complementary and not substitutive approach in which the relationship of professionals with technological tools will gain a reinforced protagonism in a context with a greater presence of semi-automated content. Content monitoring and error checking (**DalBen**; **Jurno**, 2021), or the development of interpretive and opinionated texts not based on existing data, may be specific roles for journalists, among whom the need for professional control over automated content seems to have become normalized (**Wu**; **Tandoc**; **Salmon**, 2019).

Finally, audiences are perceived as the fundamental element in the appropriation and thus success of AI in media (**Kim**; **Kim**, 2017), and the way in which automated content has been received so far indicates that there is no negative predisposition to its implementation. The perceived credibility of this content is very similar to that attributed to journalists, and the differences in the perceived quality of the texts are also small. Aspects of enjoyableness and readability –particularly in texts written in a nonobjective way- are, so far, the main differential elements in favor of journalists.

It should be noted that the automated texts existing to date have focused on relatively peripheral and highly structured thematic areas associated with reliable databases, such as weather or stock market information. Outside such domains, and particularly in more interpretive texts, it should be noted how advances in generative AI can broaden the acceptability of the produced texts (**Graefe** *et al.*, 2018).

To the extent that AI and data intelligence have the potential to become a regular resource for major content producers and media companies in the industry in the medium and long term, an approach is required that goes beyond the purely technological and addresses challenges such as quality and transparency, respect for privacy, the fight against information disruption or social development.

A change in the focus of the debate is thus encouraged: to not simply talk about what AI is capable of doing, but rather to focus on analyzing how it can affect people and journalists, how it can be used for the proper purposes of the profession and social good, and how to close the gaps that its use can cause (**Broussard** *et al.*, 2019; **Riedl**, 2019; **Deuze**; **Beckett**, 2022).

## 6. References

**Adami, Marina** (2023). “Is ChatGPT a threat or an opportunity for journalism? Five AI experts weigh in”. Oxford: Reuters Institute for the Study of Journalism.
*https://reutersinstitute.politics.ox.ac.uk/news/chatgpt-threat-or-opportunity-journalism-five-ai-experts-weigh*

**Ali, Waleed**; **Hassoun, Mohamed** (2019). “Artificial intelligence and automated journalism: Contemporary challenges and new opportunities”. *International journal of media, journalism and mass communications*, v. 5, n. 1, pp. 40-49.
*https://doi.org/10.20431/2454-9479.0501004*

**Anderson, C. W.** (2012). “Towards a sociology of computational and algorithmic journalism”. *New media & society*, v. 15, n. 7, pp. 1005-1021.
*https://doi.org/10.1177/1461444812465137*

**Andersson, Ulrika**; **Wiik, Jenny** (2013). “Journalism meets management”. *Journalism practice*, v. 7, n. 6, pp. 705-719.
*https://doi.org/10.1080/17512786.2013.790612*

**Barceló-Ugarte, Teresa**; **Pérez-Tornero, José-Manuel**; **Vila-Fumàs, Pere** (2021). “Ethical challenges in incorporating artificial intelligence into newsrooms”. In: *News media innovation reconsidered*. M. Luengo; S. Herrera-Damas (eds.).
*https://doi.org/10.1002/9781119706519.ch9*

**Beckett, Charlie** (2019). *New powers, new responsibilities: A global survey of journalism and artificial intelligence*. The London School of Economics and Political Science.
*https://blogs.lse.ac.uk/polis/2019/11/18/new-powers-new-responsibilities*

**Biswal, Santosh-Kumar**; **Gouda, Nikhil-Kumar** (2020). “Artificial intelligence in journalism: a boon or bane?”. In: Kulkarni, A.; Satapathy, S. (eds.). *Optimization in machine learning and applications. algorithms for intelligent systems*. Singapore: Springer.
*https://doi.org/10.1007/978-981-15-0994-0_10*

**Boczkowsi, Pablo** (2004). *Digitizing the news*. Cambridge: MIT Press.

**Brennen, J. Scott**; **Howard, Philip N.**; **Nielsen, Rasmus-Kleis** (2018). *An industry-led debate: how UK media cover artificial intelligence*. Reuters Institute for the Study of Journalism.
*https://www.oxfordmartin.ox.ac.uk/publications/an-industry-led-debate-how-uk-media-cover-artificial-intelligence*

**Brennen, J. Scott**; **Howard, Philip N.**; **Nielsen, Rasmus-Kleis** (2022). "What to expect when you're expecting robots: Futures, expectations, and pseudo-artificial general intelligence in UK news". *Journalism*, v. 23, n. 1, pp. 22-38.
*https://doi.org/10.1177/1464884920947535*

**Broussard, Meredith**; **Diakopoulos, Nicholas**; **Guzman, Andrea**; **Abebe, Rediet**; **Dupagne, Michel**; **Chuan, Ching-Hua** (2019). "Artificial intelligence and journalism". *Journalism and mass communication quarterly*, v. 96, n. 3, pp. 673-695.
*http://doi.org/10.1177/1077699019859901*

**Brundage, Miles**; **Avin, Shahar**; **Clark, Jack**; **Toner, Helen**; **Eckersley, Peter**; **Garfinkel, Ben**; **Dafoe, Allan**; **Scharre, Paul**; **Zeitzoff, Thomas**; **Filar, Bobby**; **Anderson, Hyrum**; **Roff, Heather**; **Allen, Gregory C.**; **Steinhardt, Jacob**; **Flynn, Carrick**; **Héigeartaigh, Seán Ó.**; **Beard, Simon**; **Belfield, Haydn**; **Farquhar, Sebastian**; **Lyle, Clare**; **Crootof, Rebecca**; **Evans, Owain**; **Page, Michael**; **Bryson, Joanna**; **Yampolskiy, Roman**; **Amodei, Dario** (2018). *The malicious use of artificial intelligence: Forecasting, prevention, and mitigation*.
*https://arxiv.org/ftp/arxiv/papers/1802/1802.07228.pdf*

**Calvo-Rubio, Luis-Mauricio**; **Ufarte-Ruiz, María-José** (2020). "Percepción de docentes universitarios, estudiantes, responsables de innovación y periodistas sobre el uso de inteligencia artificial en periodismo". *Profesional de la información,* v. 29, n. 1.
*https://doi.org/10.3145/epi.2020.ene.09*

**Calvo-Rubio, Luis-Mauricio**; **Ufarte-Ruiz, María-José** (2021). "Inteligencia artificial y periodismo: Revisión sistemática de la producción científica en Web of Science y Scopus (2008-2019)". *Communication & society*, v. 34, n. 2, pp. 159-176.
*https://doi.org/10.15581/003.34.2.159-176*

**Canavilhas, João** (2022). "Artificial intelligence and journalism: Current situation and expectations in the Portuguese sports media". *Journalism and media*, v. 3, n. 3, pp. 510-520.
*https://doi.org/10.3390/journalmedia3030035*

**Carlson, Matt** (2015). "The robotic reporter". *Digital journalism*, v. 3, n. 3, pp. 416-431.
*https://doi.org/10.1080/21670811.2014.976412*

**Chadwick, Andrew** (2013). *Hybrid media system: politics and power*. Oxford Studies in Digital Politics.
*https://doi.org/10.1093/acprof:oso/9780199759477.001.0001*

**Chan-Olmsted, Sylvia M.** (2019). "A review of artificial intelligence adoptions in the media industry". *International journal on media management*, v. 21, n. 3-4, pp. 193-215.
*https://doi.org/10.1080/14241277.2019.1695619*

**Clerwall, Christer** (2014). "Enter the robot journalist". *Journalism practice*, v. 8, n. 5, pp. 519-531.
*http://doi.org/10.1080/17512786.2014.883116*

**Codina, Lluís** (2020). "Revisiones bibliográficas sistematizadas en Ciencias Humanas y Sociales. 1: Fundamentos". En: Lopezosa, C., Díaz-Noci, J., Codina, L. (ed.). *Methodos, Anuario de métodos de investigación en comunicación social*, 1 (pp. 50-60). Barcelona: Universitat Pompeu Fabra.
*https://doi.org/10.31009/methodos.2020.i01.05*

**Codina Lluís**; **Vállez, Mari** (2018). Periodismo computacional: evolución, casos y herramientas. *Profesional de la información*, v. 27, n. 4, pp. 759-768.
*https://doi.org/10.3145/epi.2018.jul.05*

*Comisión Europea* (2020). *Libro blanco sobre la inteligencia artificial. Un enfoque europeo orientado a la excelencia y la confianza.*
*https://op.europa.eu/es/publication-detail/-/publication/ac957f13-53c6-11ea-aece-01aa75ed71a1*

**Cools, Hannes**; **Van-Gorp, Baldwin**; **Opgenhaffen, Michael** (2022). "Where exactly between utopia and dystopia? A framing analysis of AI and automation in US newspapers". *Journalism*.
*https://doi.org/10.1177/14648849221122647*

**DalBen, Silvia**; **Jurno, Amanda** (2021). "More than code: The complex network that involves journalism production in five Brazilian robot initiatives". *#ISOJ Journal*, v. 11, n. 1, pp. 111-137.
*https://isoj.org/research/more-than-code-the-complex-network-that-involves-journalism-production-in-five-brazilian-robot-initiatives*

**Danzon-Chambaud, Samuel** (2021). "A systematic review of automated journalism scholarship: guidelines and suggestions for future research". *Open research Europe*, v. 1, n. 4.
*https://doi.org/10.12688/openreseurope.13096.1*

**De-Haan, Yael**; **Van-den-Berg, Eric**; **Goutier, Nele**; **Kruikemeier, Sanne**; **Lecheler, Sophie** (2022). "Invisible friend or foe? How journalists use and perceive algorithmic-driven tools in their research process". *Digital journalism*.
*https://doi.org/10.1080/21670811.2022.2027798*

**De-Lara-González, Alicia**; **Arias-Robles, Félix**; **Carvajal-Prieto, Miguel**; **García-Avilés, José-Alberto** (2015). "Ranking de innovación periodística 2014 en España. Selección y análisis de 25 iniciativas". *Profesional de la información*, v. 24, n. 3, pp. 235-245.
*https://doi.org/10.3145/epi.2015.may.03*

**De-Lara-González, Alicia**; **García-Avilés, José-Alberto**; **Arias-Robles, Félix** (2022). "Implantación de la Inteligencia Artificial en los medios españoles: análisis de las percepciones de los profesionales". *Textual & visual media*, v. 1, n. 15.
*https://doi.org/10.56418/txt.15.2022.001*

**De-Lima-Santos, Mathias-Felipe**; **Ceron, Wilson** (2021). "Artificial intelligence in news media: current perceptions and future outlook". *Journalism and media*, v. 3, n. 1, pp. 13-26.
*https://doi.org/10.3390/journalmedia3010002*

**De-Lima-Santos, Mathias-Felipe**; **Salaverría, Ramón** (2021). "Del periodismo de datos a la inteligencia artificial: desafíos que enfrenta La Nación en la implementación de la visión artificial para la producción de noticias". *Palabra clave*, v. 24, n. 3, e2437.
*https://doi.org/10.5294/pacla.2021.24.3.7*

**Deuze, Mark** (2005). "What is journalism? Professional identity and ideology of journalists reconsidered". *Journalism*, v. 6, n. 4, pp. 442-464.
*https://doi.org/10.1177/1464884905056815*

**Deuze, Mark** (2009). "Media industries, work and life". *European journal of communication*, v. 24, n. 4.
*https://doi.org/10.1177/0267323109345523*

**Deuze, Mark**; **Beckett, Charlie** (2022). "Imagination, algorithms and news: Developing AI literacy for journalism". *Digital journalism*.
*https://doi.org/10.1080/21670811.2022.2119152*

**Diakopoulos, Nicholas** (2019). *Automating the news: How algorithms are rewriting the media*. Cambridge, MA: Harvard University Press.
*https://www.jstor.org/stable/j.ctv24w634d*

**Díaz-Noci, Javier** (2020). "Artificial intelligence systems-aided news and copyright: Assessing legal implications for journalism practices". *Future internet*, v. 12, n. 5.
*https://doi.org/10.3390/fi12050085*

**Díaz-Noci, Javier** (2023). *Investigar la brecha digital, las noticias y los medios: hacia la equidad informativa digital*. Barcelona: Ediciones Profesionales de la Información.
*https://doi.org/10.3145/digidoc-informe8*

**Dörr, Konstantin-Nicholas** (2015). "Mapping the field of algorithmic journalism". *Digital journalism*, v. 4, n. 6, pp. 700-722.
*https://doi.org/10.1080/21670811.2015.1096748*

**Dörr, Konstantin-Nicholas**; **Hollnbuchner, Katharina** (2017). "Ethical challenges of algorithmic journalism". *Digital journalism*, v. 5, n. 4, pp. 404-419.
*https://doi.org/10.1080/21670811.2016.1167612*

**Echeverría, Javier** (2008). "El Manual de Oslo y la innovación social". *Arbor*, v. 184, n. 732, pp. 609-618.
*https://doi.org/10.3989/arbor.2008.i732.210*

**Eloundou, Tyna**; **Manning, Sam**; **Mishkin, Pamela**; **Rock, Daniel** (2023). "GPTs are GPTs: An early look at the labor market impact potential of large language models". *arXiv*.
*https://doi.org/10.48550/arXiv.2303.10130*

*European Commission* (2022a). *Communication from the Commission to the European Parliament, the Council, the European Economic and Social Committee and the Committee of the Regions on the European democracy action plan*. Brussels: European Commission.
*https://eur-lex.europa.eu/legal-content/EN/TXT/?uri=COM:2020:790:FIN*

*European Commission* (2022b). *The strengthened code of practice on disinformation*. Brussels: European Commission.
*https://digital-strategy.ec.europa.eu/en/library/2022-strengthened-code-practice-disinformation*

**Eubanks, Virginia** (2017). *Automating inequality*. New York: St Martin's Press. ISBN: 978 1 250074317

**Fanta, Alexander** (2017). *Putting Europe's robots on the map: Automated journalism in news agencies*. Reuters Institute for the Study of Journalism.
*https://reutersinstitute.politics.ox.ac.uk/sites/default/files/2017-09/Fanta%2C%20Putting%20Europe%E2%80%99s%20Robots%20on%20the%20Map.pdf*

*FAPE* (2014). *Manifiesto #LibertadDePrensa: Sin periodistas, no hay periodismo y sin periodismo, no hay democracia*. 29 de abril.
*https://fape.es/manifiesto-libertaddeprensa-sin-periodistas-no-hay-periodismo-y-sin-periodismo-no-hay-democracia*

**Ferrucci, Patrick**; **Vos, Tim** (2017). "Who's in, who's out?". *Digital journalism*, v. 5, n. 7, pp. 868-883.
*https://doi.org/10.1080/21670811.2016.1208054*

**Flew, Terry**; **Spurgeon, Christina**; **Daniel, Anna**; **Swift, Adam** (2012). "The promise of computational journalism". *Journalism practice*, v. 6, n. 2, pp. 157-171.
*https://doi.org/10.1080/17512786.2011.616655*

**García-Marín, David** (2022). "Modelos algorítmicos y fact-checking automatizado. Revisión sistemática de la literatura". *Documentación de las ciencias de la información*, v. 45, n. 1, pp. 7-16.
*https://doi.org/10.5209/dcin.77472*

**García-Orosa, Berta**; **Canavilhas, João**; **Vázquez-Herrero, Jorge** (2023). "Algorithms and communication: A systematized literature review". *Comunicar*, v. 74, pp. 9-21.
*https://doi.org/10.3916/C74-2023-01*

**Gillespie, Tarleton** (2014). "The relevance of algorithms". In: Gillespie, T.; Boczkowski, P. J.; Foot, K. A. (eds.). *Media technologies: Essays on communication, materiality, and society*. Cambridge: The MIT Press, pp. 131-150.
*https://doi.org/10.7551/mitpress/9780262525374.003.0009*

**Graefe, Andreas** (2016). *Guide to automated journalism*. Tow Center for Digital Journalism.
*https://doi.org/10.7916/D80G3XDJ*

**Graefe, Andreas**; **Bohlken, Nina** (2020). "Automated journalism: A meta-analysis of readers' perceptions of human-written in comparison to automated news". *Media and communication*, v. 8, n. 3, pp. 50-59.
*https://doi.org/10.17645/mac.v8i3.3019*

**Graefe, Andreas**; **Haim, Mario**; **Haarmann, Bastian**; **Brosius, Hans-Bernd** (2018). "Readers' perception of computer-generated news: Credibility, expertise, and readability". *Journalism*, v. 19, n. 5, pp. 595-610.
*https://doi.org/10.1177/1464884916641269*

**Grant, Maria J.**; **Booth, Andrew** (2009). "A typology of reviews: an analysis of 14 review types and associated methodologies". *Health information & libraries journal*, v. 26, pp. 91-108.
*https://doi.org/10.1111/j.1471-1842.2009.00848.x*

**Guéroult, Marcial** (1966). *El concepto de la información en la ciencia contemporánea*. Madrid: Siglo XXI Editores. ISBN: 9682304857

**Guzman Andrea L.**; **Lewis Seth C.** (2020). "Artificial intelligence and communication: A human-machine communication research agenda". *New media & society*, v. 22, n. 1.
*https://doi.org/10.1177/1461444819858691*

**Gynnild, Astrid** (2014). "Journalism innovation leads to innovation journalism: The impact of computational exploration on changing mindsets". *Journalism*, v. 15, n. 6, pp. 713-730.
*https://doi.org/10.1177/1464884913486393*

**Haim, Mario**; **Graefe, Andreas** (2017). "Automated news: Better tan expected?". *Digital journalism*, v. 5, n. 8, pp. 1044-1059.
*https://doi.org/10.1080/21670811.2017.1345643*

**Hansen, Mark**; **Roca-Sales, Meritxell**; **Keegan, Jonathan**; **King, George** (2017). *Artificial intelligence: Practice and implications for journalism*. Columbia Journalism School.
*https://doi.org/10.7916/D8X92PRD*

**Holt, Kristoffer**; **Karlsson, Michael** (2014). "'Random acts of journalism?' How citizen journalists tell the news in Sweden". *New media & society*, v. 17, n. 11, pp. 1795-1810.
*https://doi.org/10.1177/1461444814535189*

**Joris, Glen**; **De-Grove, Frederick**; **Van-Damme, Kristin**; **De-Marez, Lieven** (2021). "Appreciating news algorithms: Examining audiences' perceptions to different news selection mechanisms". *Digital journalism*, v. 9, n. 5, pp. 589-618.
*https://doi.org/10.1080/21670811.2021.1912626*

**Kapuściński, Ryszard** (2002). *Los cínicos no sirven para este oficio. Sobre el buen periodismo*. Madrid: Anagrama. ISBN: 8433967967

**Karnouskos, Stamatis** (2020). "Artificial intelligence in digital media: The era of deepfakes". *IEEE Transactions on technology and society*, v. 1, n. 3, pp. 138-147.
*https://doi.org/10.1109/TTS.2020.3001312*

**Kim, Daewon**; **Kim, Seongcheol** (2017). "Newspaper companies' determinants in adopting robot journalism". *Technological forecasting and social change*, v. 117, pp. 184-195.
*https://doi.org/10.1016/j.techfore.2016.12.002*

**Kim, Daewon**; **Kim, Seongcheol** (2018). "Newspaper journalists' attitudes towards robot journalism". *Telematics and informatics*, v. 35, n. 2, pp. 340-357.
*https://doi.org/10.1016/j.tele.2017.12.009*

**Jones, Bronwyn**; **Jones, Rhianne** (2019). "Public service chatbots: Automating conversation with BBC News". *Digital journalism*, v. 7, n. 8, pp. 1032-1053.
*https://doi.org/10.1080/21670811.2019.1609371*

**Jung, Jaemin**; **Song, Haeyeop**; **Kim, Youngju**; **Im, Hyunsuk**; **Oh, Sewook** (2017). "Intrusion of software robots into journalism: The public's and journalists' perceptions of news written by algorithms and human journalists". *Computers in human behavior*, v. 71, pp. 291-298.
*https://doi.org/10.1016/j.chb.2017.02.022*

**Lee, Jaehun**; **Suh, Taewon**; **Roy, Daniel**; **Baucus, Melissa** (2019). "Emerging technology and business model Innovation: The case of Artificial Intelligence". *Journal of open innovation: Technology, market, and complexity*, v. 5, n. 44.
*https://doi.org/10.3390/joitmc5030044*

**Lemelshtrich-Latar, Noam**; **Nordfors, David** (2009). "Digital identities and journalism content-how artificial intelligence and journalism may co-develop and why society should care". *Innovation journalism*, v. 6, n. 7, pp. 3-47.
*http://journal.innovationjournalism.org/2009/11/digital-identities-and-journalism.html*

**Lewis, Seth C.**; **Sanders, Amy-Kristin**; **Carmody, Casey** (2018). "Libel by algorithm? Automated journalism and the threat of legal liability". *Journalism & mass communication quarterly*, v. 96, n. 1, pp. 60-81.
*https://doi.org/10.1177/1077699018755983*

**Lewis, Seth C.**; **Guzman, Andrea L.**; **Schmidt, Thomas R.** (2019). "Automation, journalism, and human-machine communication: Rethinking roles and relationships of humans and machines in news". *Digital journalism*, v. 7, n. 4, pp. 409-427.
*https://doi.org/10.1080/21670811.2019.1577147*

**Lin, Bibo**; **Lewis, Seth C.** (2022). "The one thing journalistic AI just might do for democracy". *Digital journalism*, v. 10, n. 10, pp. 1627-1649.
*https://doi.org/10.1080/21670811.2022.2084131*

**Lindén, Carl-Gustav** (2017). "Decades of automation in the newsroom". *Digital journalism*, v. 5, n. 2, pp. 123-140.
*https://doi.org/10.1080/21670811.2016.1160791*

**Lindén, Carl-Gustav**; **Tuulonen, Hanna** (2019). *News automation: The rewards, risks and realities of 'machine journalism'*. WAN-IFRA report.
*https://jyx.jyu.fi/bitstream/handle/123456789/67003/2/WAN-IFRA_News_Automation-FINAL.pdf*

**Liu, Bingjie**; **Wei, Lewen** (2019). "Machine authorship in situ". *Digital journalism*, v. 7, n. 5, pp. 635-657.
*https://doi.org/10.1080/21670811.2018.1510740*

**Lokot, Tetyana**; **Diakopoulos, Nicholas** (2016). "News bots". *Digital journalism*, v. 4, n. 6, pp. 682-699.
*https://doi.org/10.1080/21670811.2015.1081822*

**López-Jiménez, Eduardo-Alejandro**; **Ouariachi, Tania** (2020). "An exploration of the impact of artificial intelligence (AI) and automation for communication professionals". *Journal of information, communication and ethics in society*, v. 19, n. 2, pp. 249-267.
*https://doi.org/10.1108/JICES-03-2020-0034*

**Marconi, Francesco** (2020). *Newsmakers: Artificial intelligence and the future of journalism*. New York: Columbia University Press. ISBN: 978 0 231549356

**Masip, Pere**; **Ruiz-Caballero, Carlos**; **Suau, Jaume** (2019). "Audiencias activas y discusión social en la esfera pública digital. Artículo de revisión. *Profesional de la información*, v. 28, n. 2.
*https://doi.org//10.3145/epi.2019.mar.04*

**Masip, Pere**; **Suau, Jaume**; **Ruiz-Caballero, Carlos** (2020). "Percepciones sobre medios de comunicación y desinformación: ideología y polarización en el sistema mediático español". *Profesional de la información*, v. 29, n 5.
*https://doi.org/10.3145/epi.2020.sep.27*

**Melin, Magnus**; **Bäck, Asta**; **Södergård, Caj**; **Munezero, Myriam D.**; **Leppänen, Leo J.**; **Toivonen, Hannu** (2018). "No landslide for the human journalist - An empirical study of computer-generated election news in Finland". *IEEE Access*, v. 6, pp. 43356-43367.
*https://doi.org/10.1109/ACCESS.2018.2861987*

**Micó, Josep-Lluís**; **Casero-Ripollés, Andreu**; **García-Orosa, Berta** (2022). "Platforms in journalism 4.0: The impact of the fourth industrial revolution on the news industry". In: Vázquez-Herrero, J.; Silva-Rodríguez, A.; Negreira-Rey, M. C.; Toural-Bran, C.; López-García, X. (eds.). *Total journalism. Studies in big data*, v. 97. Cham: Springer.
*https://doi.org/10.1007/978-3-030-88028-6_18*

**Møller, Lynge-Asbjørn** (2022). "Between personal and public interest: How algorithmic news recommendation reconciles with journalism as an ideology". *Digital journalism*, v. 10, n. 10, pp. 1794-1812.
*https://doi.org/10.1080/21670811.2022.2032782*

**Montal, Tal**; **Reich, Zvi** (2017). "I, robot. You, journalist. Who is the author?". *Digital journalism*, v. 5, n. 7, pp. 829-849.
*https://doi.org/10.1080/21670811.2016.1209083*

**Moran, Rachel E.**; **Shaikh, Sonia-Jawaid** (2022). "Robots in the news and newsrooms: Unpacking meta-journalistic discourse on the use of artificial intelligence in journalism". *Digital journalism*, v. 10, n. 10, pp. 1756-1774.
*https://doi.org/10.1080/21670811.2022.2085129*

**Nielsen, Rasmus-Kleis**; **Ganter, Sarah-Anne** (2022). *The power of platforms shaping media and society*. Oxford Studies in Digital Politics.
*https://reutersinstitute.politics.ox.ac.uk/news/power-platforms*

**Noain-Sánchez, Amaya** (2022). "Addressing the impact of artificial intelligence on journalism: The perception of experts, journalists and academics". *Communication & society*, v. 35, n. 3.
*https://doi.org/10.15581/003.35.3.105-121*

**Örnebring, Henry** (2013). "Anything you can do, I can do better? Professional journalists on citizen journalism in six European countries". *International communication gazette*, v. 75, n. 1, pp. 35-53.
*https://doi.org/10.1177/1748048512461761*

**Parratt-Fernández, Sonia**; **Mayoral-Sánchez, Javier**; **Mera-Fernández, Montse** (2021). "The application of artificial intelligence to journalism: An analysis of academic production". *Profesional de la información*, v. 30, n. 3.
*https://doi.org/10.3145/epi.2021.may.17*

**Paulussen, Steve** (2016). "Innovation in the newsroom". In: *The SAGE handbook of digital journalism*. Chapter 13. ISBN: 978 1 473906532
*https://doi.org/10.4135/9781473957909*

**Peña-Fernández, Simón**; **Lazkano-Arrillaga, Iñaki**; **García-González, Daniel** (2016). "La transición digital de los diarios europeos: Nuevos productos y nuevas audiencias". *Comunicar*, v. 46, pp. 27-36.
*https://doi.org/10.3916/C46-2016-03*

**Peña-Fernández, Simón**; **Lazkano-Arrillaga, Iñaki**; **Larrondo-Ureta, Ainara** (2019). "Medios de comunicación e innovación social. El auge de las audiencias activas en el entorno digital". *Andamios*, v. 16, n. 40, pp. 351-372.
*https://doi.org/10.29092/uacm.v16i40.710*

**Pérez-Dasilva, Jesús-Ángel**; **Mendiguren-Galdospin, Terese**; **Meso-Ayerdi, Koldobika**; **Larrondo-Ureta, Ainara**; **Peña-Fernández, Simón**; **Ganzabal-Learreta, M.**; **Lazkano-Arrillaga, I.** (2021). *Perfiles digitales de los periodistas vascos y diálogo con las audiencias*. Universidad del País Vasco.

**Pihlajarinne, Taina**; **Alén-Savikko, Anette** (2022). *Artificial intelligence and the media. Reconsidering rights and responsibilities*. Cheltenham, UK: Edward Elgar Publishing.
*https://www.elgaronline.com/view/edcoll/9781839109966/9781839109966.xml*

**Powers, Matthew** (2012). "'In forms that are familiar and yet-to-be invented': American journalism and the discourse of technologically specific work". *Journal of communication inquiry*, v. 36, n. 1, pp. 24-43.
*https://doi.org/10.1177/0196859911426009*

**Primo, Alex**; **Zago, Gabriela** (2015). "Who and what do journalism?". *Digital journalism*, v. 3, n. 1, pp. 38-52.
*https://doi.org/10.1080/21670811.2014.927987*

**Riedl, Mark O.** (2019). "Human-centered artificial intelligence and machine learning". *Human behavior and emerging technologies*, v. 1, n. 1, pp. 33-36.
*https://doi.org/10.48550/arXiv.1901.11184*

**Rojas-Torrijos, José-Luis** (2019). "Automated sports coverages. Case study of bot released by The Washington Post during Río 2016 and Pyeongchang 2018 Olympics". *Revista latina de comunicación social*, v. 74, pp. 1729-1747.
*https://doi.org/10.4185/RLCS-2019-1407*

**Ruffo, Giancarlo**; **Semeraro, Alfonso** (2022). "FakeNewsLab: Experimental study on biases and pitfalls preventing us from distinguishing true from false news". *Future internet*, v. 14, pp. 283-207.
*https://doi.org/10.3390/fi14100283*

**Sánchez-García, Pilar**; **Merayo-Álvarez, Noemí**; **Calvo-Barbero, Carla**; **Díez-Gracia, Alba** (2023). "Spanish technological development of artificial intelligence applied to journalism: companies and tools for documentation, production and distribution of information". *Profesional de la información*, v. 32, n. 2.
*https://doi.org/10.3145/epi.2023.mar.08*

**Simon, Felix M.** (2022). "Uneasy bedfellows: AI in the news, platform companies and the issue of journalistic autonomy". *Digital journalism*, v. 10, n. 10, pp. 1832-1854.
*https://doi.org/10.1080/21670811.2022.2063150*

**Shilton, Katie** (2018). "Values and ethics in human-computer interaction". *Foundations and trends in human-computer interaction*, v. 12, n. 2, pp. 107-171.
*https://doi.org/10.1561/1100000073*

**Soto-Sanfiel, María-Teresa**; **Ibiti, Adriana**; **Machado, Mabel**; **Marín-Ochoa, Beatriz-Elena**; **Mendoza-Michilot, María**; **Rosell-Arce, Claudio-Guillermo**; **Angulo-Brunet, Ariadna** (2022). "In search of the Global South: assessing attitudes of Latin American journalists to artificial intelligence in journalism". *Journalism studies*, v. 23, n. 10, pp. 1197-1224.
*https://doi.org/10.1080/1461670X.2022.2075786*

**Stenbom, Agnes**; **Wiggberg, Mattias**; **Norlund, Tobias** (2021). "Exploring communicative AI: Reflections from a Swedish newsroom". *Digital journalism*.
*https://doi.org/10.1080/21670811.2021.2007781*

**Stray, Jonathan** (2019). "Making artificial intelligence work for investigative journalism". *Digital journalism*, 7, n. 8, pp. 1076-1097.
*https://doi.org/10.1080/21670811.2019.1630289*

**Strömbäck, Jesper** (2005). "In search of a standard: four models of democracy and their normative implications for Journalism". *Journalism studies*, v. 6, n. 3, pp. 331-345.
*https://doi.org/10.1080/14616700500131950*

**Suárez-Villegas, Juan-Carlos** (2017). "Citizen journalism. Analysis of opinions of journalists from Spain, Italy and Belgium". *Convergencia*, v. 24, n. 74, pp. 91-111.
*https://doi.org/10.29101/crcs.v0i74.4383*

**Sundar, S. Shyam** (1999). "Exploring receivers' criteria for perception of print and online news". *Journalism & mass communication quarterly*, v. 76, n. 2, pp. 373-386.
*https://doi.org/10.1177/107769909907600213*

**Sundar, S. Shyam** (2008). "The MAIN model: A heuristic approach to understanding technology effects on credibility". In: *Digital media, youth, and credibility*. Miriam J. Metzger; Andrew J. Flanagin (eds.), pp. 73-100. Cambridge, MA: The MIT Press.

**Sundar, S. Shyam**; **Waddell, T. Franklin**; **Jung, Eun-Hwa** (2016). "The Hollywood robot syndrome: Media effects on older adults' robot attitudes and adoption intentions". *Proceedings of HRI 2016: ACM/IEEE International conference on human-robot interaction*, pp. 343-350.
*https://doi.org/10.1109/HRI.2016.7451771*

**Tandoc, Edson C.**; **Yao, L. Y.**; **Wu, S.** (2020). "Man vs. machine? The impact of algorithm authorship on news credibility. *Digital journalism*, v. 8, n. 4, pp. 548-562.
*https://doi.org/10.1080/21670811.2020.1762102*

**Tejedor, Santiago**; **Vila, Pere** (2021). "Exo journalism: A conceptual approach to a hybrid formula between journalism and artificial intelligence". *Journalism and media*, v. 2, n. 4.
*https://doi.org/10.3390/journalmedia2040048*

**Thurman, Neil**; **Dörr, Konstantin**; **Kunert, Jessica** (2017). "When reporters get hands-on with robo-writing". *Digital journalism*, v. 5, n. 10, pp. 1240-1259.
*https://doi.org/10.1080/21670811.2017.1289819*

**Túñez-López, José-Miguel**; **Toural-Bran, Carlos**; **Cacheiro-Requeijo, Santiago** (2018). "Uso de bots y algoritmos para automatizar la redacción de noticias: percepción y actitudes de los periodistas en España". *Profesional de la información*, v. 27, n. 4, pp. 750-758.
*https://doi.org/10.3145/epi.2018.jul.04*

**Túñez-López, José-Miguel**; **Toural-Bran, Carlos**; **Valdiviezo-Abad, Cesibel** (2019). "Automation, bots and algorithms in newsmaking. Impact and quality of artificial journalism". *Revista latina de comunicación social*, v. 74.
*https://minerva.usc.es/xmlui/bitstream/handle/10347/24522/2019_rlcs_tunez_automatizacion_eng.pdf*

**Ufarte-Ruiz, María-José**; **Calvo-Rubio, Luis-Mauricio**; **Murcia-Verdú, Francisco-José** (2021). "Los desafíos éticos del periodismo en la era de la inteligencia artificial". *Estudios sobre el mensaje periodístico*, v. 27, n. 2, pp. 673-684.
*https://doi.org/10.5209/esmp.69708*

**Ufarte-Ruiz, María-José**; **Murcia-Verdú**, **Francisco-José**; **Túñez-López, José-Miguel** (2023). "Use of artificial intelligence in synthetic media: first newsrooms without journalists". *Profesional de la información*, v. 32, n. 2.
*https://doi.org/10.3145/epi.2023.mar.03*

**Van-Dalen, Arjen** (2012). "The algorithms behind the headlines". *Journalism practice*, v. 6, n. 5-6, pp. 648-658.
*https://doi.org/10.1080/17512786.2012.667268*

**Van-Drunen, Max Z.**; **Fechner, D.** (2022). "Safeguarding editorial independence in an automated media system: The relationship between law and journalistic perspectives". *Digital journalism*.
*https://doi.org/10.1080/21670811.2022.2108868*

**Van-Lente, Harro** (2012). "Navigating foresight in a sea of expectations: Lessons from the sociology of expectations". *Technology analysis & strategic management*, v. 24, pp. 769-782.
*https://doi.org/10.1080/09537325.2012.715478*

**Veglis, Andreas**; **Maniou, Theodora A.** (2019). "Chatbots on the rise: A new narrative in journalism". *Studies in media and communication*, v. 7, n. 1.
*https://doi.org/10.11114/smc.v7i1.3986*

**Ventura-Pocino, Patrícia** (2021). *Algorithms in the newsrooms. Challenges and recomendations for artificial intelligence with the ethical values of journalism*. Barcelona: Catalan Press Council.
*https://fcic.periodistes.cat/wp-content/uploads/2022/03/venglishDIGITAL_ALGORITMES-A-LES-REDACCIONS_ENG-1.pdf*

**Vos, Tim P.**; **Ferrucci, Patrick** (2018). "Who am I? Perceptions of digital journalists' professional identity". In: S. Eldridge; B. Franklin (eds.). *The Routledge handbook of developments in digital journalism studies.* Londres: Routledge. ISBN: 978 1 138283053
*https://doi.org/10.4324/9781315270449-4*

**Waddell, T. Franklin** (2018). "A robot wrote this?". *Digital journalism*, v. 6, n. 2, pp. 236-255.
*https://doi.org/10.1080/21670811.2017.1384319*

**Waddell, T. Franklin** (2019). "Can an algorithm reduce the perceived bias of news? Testing the effect of machine attribution on news readers' evaluations of bias, anthropomorphism, and credibility". *Journalism & mass communication quarterly*, v. 96, n. 1, pp. 82-100.
*https://doi.org/10.1177/1077699018815891*

**Wölker, Anja**; **Powell, Thomas E.** (2018). "Algorithms in the newsroom? News readers' perceived credibility and selection of automated journalism". *Journalism*, v. 22, n. 1, pp. 86-103.
*https://doi.org/10.1177/1464884918757072*

**Wu, Shangyuan**; **Tandoc, Edson C.**; **Salmon, Charles T.** (2019). "Journalism reconfigured". *Journalism studies*, v. 20, n. 10, pp. 1440-1457.
*https://doi.org/10.1080/1461670X.2018.1521299*

**Young, Mary-Lynn**; **Hermida, Alfred** (2015). "From Mr. and Mrs. Outlier to central tendencies". *Digital journalism*, v. 3, n. 3, pp. 381-397.
*https://doi.org/10.1080/21670811.2014.976409*

**Zheng, Yue**; **Zhong, Bu**; **Yang, Fan** (2018). "When algorithms meet journalism: The user perception to automated news in a cross-cultural context". *Computers in human behavior*, v. 86, pp. 266-275.
*https://doi.org/10.1016/j.chb.2018.04.046*